\documentclass[aps,prd,onecolumn,eqsecnum,amssymb,amsmath,showpacs,a4paper,superscriptaddress,nofootinbib,longbibliography]{revtex4-2}
\usepackage{hyperref}
\usepackage{graphicx}
\usepackage{amsfonts}
\usepackage{amsmath}

\begin{document}

\title{Tidal Forces in Majumdar--Papapetrou Spacetimes}

\author{Eduardo Albacete}
 \email{eduardo.albacete@ufabc.edu.br}
 \affiliation{Centro de Ci\^encias Naturais e Humanas, Universidade Federal do ABC (UFABC), 09210-170 Santo Andr\'e, S\~ao Paulo, Brazil}
 \affiliation{Centro de Matem\'atica, Computa\c c\~ao e Cogni\c c\~ao, Universidade Federal do ABC (UFABC), 09210-170 Santo Andr\'e, S\~ao Paulo, Brazil}
\author{Maur\'icio Richartz}
 \email{mauricio.richartz@ufabc.edu.br}
\affiliation{Centro de Matem\'atica, Computa\c c\~ao e Cogni\c c\~ao, Universidade Federal do ABC (UFABC), 09210-170 Santo Andr\'e, S\~ao Paulo, Brazil}

\begin{abstract}
  Tidal disruption events occur when astrophysical objects are destroyed by black holes due to strong tidal force effects. Tidal forces have been studied in a variety of black hole spacetimes, including Reissner--Nordstr\"om and Kerr spacetimes. Despite the vast literature on the subject, tidal forces around black holes in static equilibrium have never been investigated before. The aim of this work is to fill in this gap and explore tidal forces in the Majumdar--Papapetrou spacetime describing two extremely charged binary black holes in equilibrium. We focus on tidal forces associated with radial and circular geodesics of massive neutral particles moving on the plane equidistant to the black holes. In particular, we study the behavior of the tidal forces as a function of the distance from the black holes and as a function of the energy of the geodesics. We also investigate the numerical solutions of the geodesic deviation equation for different initial conditions.
\end{abstract}

\maketitle
%%%%%%%%%%%%%%%%%%%%%%%%%%%%%%%%%%%%%%%%%%%%%%%%%%%%%%%%%%%%%%%%%%%%%%%%%%%%%%%%%%%%%%%%%%%%%%%%

\section{Introduction}
%%%
Recent observational results~\cite{paperi01,paperi02,EventHorizonTelescope:2019dse} have increased the interest of the physics community and the general public in black holes (and also in the wide range of phenomena associated with them). The possibility of detecting and studying effects like quasinormal ringing, shadows, and superradiance through radio telescopes and gravitational wave detectors has become a reality in the last decade. The prospect of new groundbreaking results of gravitational physics related to black holes and other compact objects is very promising. 

Since the first detection of gravitational waves, more than 90 events consisting of the inspiralling and merger of binary systems of compact objects have been observed by the LIGO-Virgo collaboration in its first three observing runs~\cite{LIGOScientific:2021djp}. Theoretical and numerical studies of such binaries are crucial for a greater understanding of current and future observations. Unfortunately, there are no known analytical solutions of Einstein’s equations that describe astrophysical binary systems. Nevertheless, exact and simple solutions corresponding to the metrics of non-coalescing pairs of black holes exist. One of such solutions is the Majumdar--Papapetrou (MP) metric~\cite{MAJUMDAR1947,PAPAPETROU1947} of extremely charged black holes in static equilibrium. The MP spacetime has been successfully used in the past as a toy model to investigate how the presence of a second black hole influences quasinormal ringing, shadows, Penrose energy extraction, and other typical black hole effects~\cite{paperi08,paperi09,Bohn:2014xxa,Shipley:2016omi,Assumpcao:2018bka,Shipley:2019kfq,BINI2019,Sanches:2021kye}.

Building on recent articles~\cite{paper01,paper02,paper05,Sharif:2018gzj,Shahzad:2017vwi,Hong:2020bdb,Li:2021izh,Vandeev:2021yan,Uniyal:2022ouc} that have analyzed tidal forces in black hole spacetimes (including Reissner--Nordstr\"om and Kerr), the objective of this work is to investigate tidal forces acting on neutral massive particles that move along geodesics in the spacetime of two static black holes described by the MP metric. Tidal forces arise, for instance, when one considers a body falling towards the Earth. Due to the gravitational interaction, the body suffers stretching in the direction of motion and compression in transverse directions. In particular, tidal forces in the Schwarzschild spacetime can be used to describe the orbital motion of satellites around Earth and estimate deviations in their trajectories~\cite{paperi013}.   

This work is organized as follows. In Section II, after introducing the MP spacetime for two black holes in equilibrium, we obtain the equations of motion for massive particles moving along geodesics that lie on the plane that is equidistant from the black holes. In Section III, we determine a freely falling frame, which is adapted to the geodesics discussed in Section II. In Section IV, we review the geodesic deviation equation of General Relativity and write it down explicitly for the MP metric. Using this equation, we analyze the properties of the tidal forces acting on radial and circular geodesics. In Section V, we solve the geodesic deviation equation numerically and analyze the behavior of the corresponding solutions as a function of the radial distance. Finally, in Section VI, we review the main results of our work and discuss future research directions.

\section{The Majumdar--Papapetrou Spacetime}

The MP spacetime describes a collection of maximally charged black holes that are in static equilibrium due to the balance of electromagnetic and gravitational interactions~\cite{paper10,MAJUMDAR1947,paper07}. The MP metric, in Weyl cylindrical coordinates $(t,\rho,\phi,z)$, is 
\begin{equation}
	ds^{2} =-\frac{1}{U^{2}(\rho,z)}dt^{2}+U^{2}(\rho,z)\left(d\rho^{2} + \rho^{2}d\phi^{2} +dz^{2} \right),  
	\label{eq10}
\end{equation}
where $U(\rho,z)$ denotes the associated electromagnetic potential. In this work, we focus on a binary system of equal-mass black holes and assume that the black holes are located along the z-axis, at $z=\pm b$ (the quantity $2b$, hence, measures the separation between the black holes). With these assumptions, the electromagnetic potential takes the explicit form: 
\begin{equation}
	U(\rho,z) = 1 + \frac{M}{\sqrt{\rho^{2} + (z-b)^2}} + \frac{M}{\sqrt{\rho^{2} + (z+b)^2}},
	\label{eq2}
\end{equation}
where $M$ denotes the mass of each black hole (each black hole has electric charge $Q=M$). In particular, when $b=0$, the metric describes a single extremal Reissner--Nordstr\"om black hole of mass $2M$.
In our numerical analyses and figures, we have chosen $M=1$, which corresponds to normalizing the variables and parameters with respect to the mass of the black holes.

We consider timelike geodesics that correspond to the trajectories of massive particles subjected exclusively to the gravitational interaction. Geodesics, being curves that maximize the proper time between two events in a spacetime, can be determined from the Euler--Lagrange equations for the following Lagrangian $\mathcal{L}$:   
\begin{equation}
	  \mathcal{L}=g_{\mu\nu}v^{\mu}v^{\nu}=-\frac{\dot{t}^{2}}{U^{2}}+U^{2}\left( \dot{\rho}^{2} + \rho^{2}\dot{\phi}^{2} + \dot{z}^{2} \right),
	\label{eq28}
\end{equation}
where 
\begin{equation}
v^{\mu} = \frac{d}{d\tau}(t,\rho,\phi,z) = (\dot{t},\dot{\rho},\dot{\phi},\dot{z})
\end{equation}
\begin{equation} \label{lag_norm}
\mathcal{L} = -1.
\end{equation}

Taking advantage of the fact that $t$ and $\phi$ are cyclic coordinates of the Lagrangian, one can determine two invariants of motion~\cite{Assumpcao:2018bka,paper12}: the energy $E$ and the angular momentum $L$, defined, respectively, as
	\begin{align}
		E=& \frac{1}{U^{2}}\dot{t}, \label{eq4} \\
		L=& \rho^{2}U^{2} \dot{\phi}. \label{eq12}
	\end{align}
These constants of motion can also be obtained from the fact that $\partial_t$ and $\partial_\phi$ are Killing vector fields of the spacetime associated, respectively, with the stationarity and the axisymmetry of the MP metric \eqref{eq10}. 

We restrict our analysis to timelike geodesics confined to the plane $z=0$. Note that these planar geodesics only exist when the masses of the black holes are the same (as we have assumed in this work). The restriction $z=0$, together with \eqref{lag_norm}--\eqref{eq12}, implies the following equation of motion:
\begin{equation} 
	\dot{\rho}^{2} = E^{2} - V_{\mathrm{eff}}(\rho) =  E^{2} - \frac{1}{U^{2}(\rho,0)} - \frac{L^2}{\rho^2 U^{4}(\rho,0)},
	\label{eq11}
\end{equation}
where $V_{\mathrm{eff}}$ denotes the associated effective potential. This is a first-order ordinary differential equation for $\rho$, which can be solved after one chooses an initial condition. From the solution $\rho(\tau)$ of \eqref{eq11}, one can determine $t(\tau)$ using Equation~\eqref{eq4} and $\phi(\tau)$ using Equation~\eqref{eq12} to find the complete trajectory $(t(\tau),\rho(\tau),\phi(\tau),0)$. To investigate tidal forces in the $z=0$ plane, we shall assume that the geodesics are either radial (in which case, $\dot \phi = 0$ and, consequently $L=0$) or circular (in which case, $\dot \rho = 0$ and $L \neq 0$).

Note that, according to Equation~\eqref{eq11}, radial geodesic motion on the $z$ plane, corresponding to $\phi(\tau) = \phi_0 = \mathrm{constant}$, is possible only if the absolute value of the energy is greater than a minimum value $E_{\mathrm{min}}$ given by
\begin{equation} \label{bound}
	E_{\mathrm{min}} = \frac{b}{b+2M} .
\end{equation}
In particular, if $ E_{\mathrm{min}} \le |E|<1$, the trajectory is bounded and has a turning point at 
\begin{equation}
\rho = \rho_{\mathrm{max}} = \sqrt{ \left( \frac{2ME}{1-|E|} \right)^2 - b^2}.      
\end{equation}

Additionally, according to Equation~\eqref{eq11}, circular geodesic motion on the $z$ plane, corresponding to $\rho(\tau) = \rho_0 = \mathrm{constant}$, is only possible when $E$, $L$, and $\rho_0$ satisfy the following requirements:

\begin{equation}
		\frac{\partial V_{\mathrm{eff}}}{\partial \rho} = 0  \Rightarrow  \frac{\partial U}{\partial \rho} (\rho_0,0) + \frac{L^2 U(\rho_0 ,0)}{2 L^2 \rho_0 +\rho_0 ^3 U^2(\rho_0 ,0)}=0 \label{circreq1}
\end{equation}
and
\begin{equation}
  E^2= V_{\mathrm{eff}} \Rightarrow  E^2 = \frac{1}{U^2(\rho_0 ,0)}+\frac{L^2}{\rho_0 ^2 U^4(\rho_0 ,0)} . \label{circreq2}
\end{equation}

\begin{figure}[h]
\centering
\includegraphics[width=15.5cm]{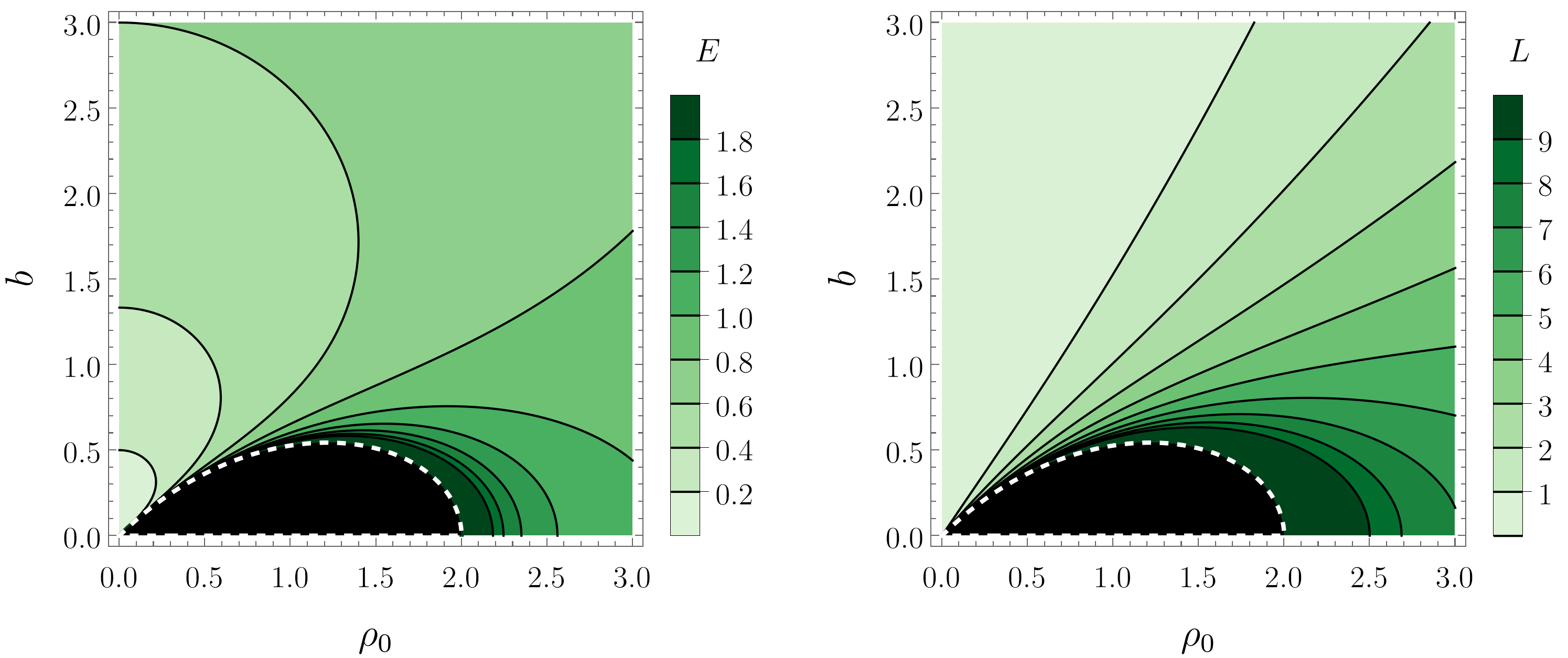}
\caption{Contour plots of the energy $E$ (\textbf{left}) and of the angular momentum $L$ (\textbf{right}) associated with circular geodesics in the $z=0$ plane, as a function of the distance $b$ between the black holes and the radius $\rho_0$ of the geodesic. The black region corresponds to the parameters for which circular geodesics are not allowed.} \label{fig0}
\end{figure}

Given $L$, one can first use \eqref{circreq1} to determine the value of $\rho_0$ corresponding to a circular geodesic. The associated energy $E$ can then be directly determined from \eqref{circreq2}. From Equation~\eqref{eq12}, we find that the angle changes linearly with respect to the proper time for circular geodesics:
\begin{equation}
\phi(\tau) = \phi_0 + \frac{L}{\rho_0^2U^2(\rho_0,0)} \tau, \label{phievol}
\end{equation}
where $\phi_0$ is the angle at $\tau=0$. In Figure \ref{fig0}, we show the values of $E$ and $L$ as a function of the radius of the circular geodesic and the separation between the black holes. We have observed that the values of $E$ and $L$ grow without bound when $b$ and $\rho_0$ approach the white dashed line in Figure \ref{fig0}. Inside the black region, Equations~\eqref{circreq1} and \eqref{circreq2} do not yield real solutions, meaning that circular geodesics are not allowed for the corresponding parameters $b$ and $\rho_0$.

\section{Freely Falling Frames in the Majumdar--Papapetrou Spacetime}
To investigate tidal forces along a geodesic, it is convenient to define a freely falling frame, i.e.,~a tetrad basis, which is adapted to the corresponding trajectory (meaning that the elements of the tetrad are parallel propagated along the geodesic). A tetrad basis $\{e_{ \hat{a}}\} = \{e _{\hat{0}},e _{\hat{1}},e _{\hat{2}},e _{\hat{3}}\}$ is a non-coordinate basis that follows a prescribed normalization~\cite{Chandrasekhar:1985kt,dInverno:1992gxs,paper13}. We denote each component (according to the coordinate basis $\partial_\mu$) of the tetrad element $e_{ \hat{a}}$ as $e^{\mu} \! _{ \hat{a}}$. The components of the corresponding dual vectors are $e _{\mu \hat{a}}$. We assume that the tetrad basis is normalized according to $e^{\mu} \! _{ \hat{a}} e _{\mu \hat{b}} =\eta_{\hat{a}\hat{b}}$, where $\eta_{\hat{a}\hat{b}} = $ \textbf{diag} $(-1, 1, 1, 1)$. We set the first element of the tetrad basis as the tangent vector of the timelike geodesic we are interested in, i.e.,~$e^{\mu} \ _{\! \! \hat{0}}=v^\mu$. The remaining elements of the tetrad basis are determined using the normalization prescribed above and {the requirement that the tetrad elements are parallel transported along the trajectory, i.e.,

\begin{equation}
\frac{D e^{\mu} \! _{ \hat{a}} }{D\tau } = 0,   
\end{equation}
where $D/D\tau = v^\mu \nabla_\mu = e^{\mu} \ _{\! \! \hat{0}}\nabla_\mu$ is the directional derivative along the geodesic. 
}

We derive below a general expression for a freely falling frame associated with geodesics confined to the plane $z=0$ of the MP spacetime. Equations~\eqref{eq4}--\eqref{eq11} determine the tangent vector of the geodesic and, therefore, also determine the first vector of the tetrad basis: 
\begin{equation} 
v^{\mu} = \lambda^{\mu} \! _{\hat{0}} = \left( EU^{2}(\rho,0), -\frac{\sqrt{E^{2}\rho^2 U^{4}(\rho,0) - \rho^2 U^{2}(\rho,0)- L^2}}{\rho U^{2}(\rho,0)}, \frac{L}{\rho^2 U^{2}(\rho,0)}, 0	\right). \label{e0mu}
\end{equation}
By definition, $\lambda^{\mu} \! _{\hat{0}}$ is parallel propagated along the geodesic. It is straightforward to find another vector, which is parallel propagated along the geodesic and which satisfies the requirements of the tetrad basis. In fact, the symmetries of the problem, together with the fact that the geodesic is confined to the $z=0$ plane, lead to: 
\begin{equation}
\lambda^{\mu} \! _{\hat{3}} = \left( 0, 0, 0, \frac{1}{U(\rho,0)} \right). \label{e3mu}
\end{equation}

The difficult part is to find two extra vectors that are parallel propagated along the geodesic. In order to do so, we follow the idea used in Ref.~\cite{marck} to investigate tidal effects in the Kerr metric. We start by determining two vectors, $\lambda^{\mu} \! _{\hat{1}}$ and $\lambda^{\mu} \! _{\hat{2}}$, which, although not parallel propagated along the geodesic, are such that $\lambda^{\mu} \! _{ \hat{a}} \lambda _{\mu \hat{b}} =\eta_{\hat{a}\hat{b}}$:
\begin{equation}
	\begin{aligned}
		\lambda^{\mu} \! _{\hat{1}} = &\left( - U \sqrt{\frac{E^{2}\rho^2 U^{4} - L^2  - \rho^2 U^{2}}{L^2 + \rho^2 U^{2}}}, \frac{E \rho U}{\sqrt{L^2 + \rho^2 U^{2}}}, 0, 0 \right), &\\
		\lambda^{\mu} \! _{\hat{2}} = &\left( \frac{E L U^2}{\sqrt{L^2 + \rho^2 U^{2}}}, - \frac{L}{\rho U^2} \sqrt{\frac{E^{2}\rho^2 U^{4} - L^2  - \rho^2 U^{2}}{L^2 + \rho^2 U^{2}}}, \frac{\sqrt{L^2 + \rho^2 U^{2}}}{\rho^2 U^2}, 0 
		\right).
	\end{aligned}
	\label{e12mu}
\end{equation}
To simplify the notation, we have omitted the dependence $(\rho,0)$ from $U(\rho,0)$ in the expression above. 

Note that one can rotate the vectors $\lambda^{\mu} \! _{\hat{1}}$ and $\lambda^{\mu} \! _{\hat{2}}$ by an arbitrary angle while maintaining the orthonormalization $\lambda^{\mu} \! _{ \hat{a}} \lambda _{\mu \hat{b}} =\eta_{\hat{a}\hat{b}}$. Using this property, we can determine under which conditions the rotated vectors become parallel transported along the geodesic. More precisely, if $\Psi=\Psi(\tau)$ denotes the rotation angle, the rotated vectors are given by  \begin{equation}
	\begin{aligned}
		e^{\mu} \! _{\hat{1}} = & \phantom{-} \lambda^{\mu} \! _{\hat{1}} \cos \Psi + \lambda^{\mu} \! _{\hat{2}} \sin \Psi, &\\
		e^{\mu} \! _{\hat{2}} = & -\lambda^{\mu} \! _{\hat{1}} \sin \Psi + \lambda^{\mu} \! _{\hat{2}} \cos \Psi.
	\end{aligned}
	\label{e12mun}
\end{equation} 
By requiring that $D(e^{\mu} \! _{\hat{1}})/D\tau = D(e^{\mu} \! _{\hat{2}})/D\tau = 0$, we find that the angle $\Psi$ must satisfy the equation
\begin{equation} \label{psidef}
\frac{d \Psi}{d \tau} = - \frac{E L}{L^2 + \rho^2 U^{2}(\rho,0)} \left( U(\rho,0) + \rho \frac{\partial U(\rho,0)}{\partial \rho}  \right).
\end{equation}
Hence, if we choose $\Psi$ according to \eqref{psidef} and define $e^{\mu} \! _{\hat{0}}=\lambda^{\mu} \! _{\hat{0}}$, $e^{\mu} \! _{\hat{3}}=\lambda^{\mu} \! _{\hat{3}}$, the tetrad basis $\{e_{ \hat{a}}\}$ will be parallel propagated along the geodesic, whose tangent vector is \eqref{e0mu}.

Note that, in general, the angle $\Psi$ will change along the geodesic. Since $E$ cannot be zero, $\Psi$ will remain constant along the geodesic only when $L=0$. In fact, if we let $L=0$ and $\Psi=0$, Equations~\eqref{e0mu}--\eqref{e12mun} reduce to:
\begin{equation}
	\begin{aligned}
		e^{\mu} \! _{\hat{0}} = &\left( EU^{2}(\rho,0), -\frac{\sqrt{E^{2}U^{2}(\rho,0)-1}}{U(\rho,0)}, 0, 0
		\right), &\\
		e^{\mu} \! _{\hat{1}} = &\left( -U(\rho,0)\sqrt{E^{2}U^{2}(\rho,0)-1}, E, 0, 0 \right), &\\
		e^{\mu} \! _{\hat{2}} = &\left( 0, 0, \frac{1}{\rho U(\rho,0)}, 0 
		\right), &\\
		e^{\mu} \! _{\hat{3}} = &\left( 0, 0, 0, \frac{1}{U(\rho,0)}  
		\right), &\\
	\end{aligned}
	\label{eq3}
\end{equation}
which, therefore, constitutes a freely falling frame for radial geodesics on the $z$ plane.

The angle $\Psi$ also takes a simple form for circular geodesics. Considering the fact that $\rho=\rho_0$ is constant along such geodesics, one finds that $d \Psi/d \tau$ is also constant. Taking into account Equations~\eqref{circreq1}--\eqref{circreq2}, one finds the following freely falling frame for circular geodesics:  
\begin{equation}
	\begin{aligned}
		e^{\mu} \! _{\hat{0}} = &\left( EU^{2}(\rho_0,0), 0, \frac{L}{\rho_0^2 U^2(\rho_0,0)}, 0
		\right), &\\
		e^{\mu} \! _{\hat{1}} = &\left( \frac{-L \sin \Psi}{\rho_0} , \frac{\cos \Psi}{U(\rho_0,0)} , -\frac{E \sin \Psi}{\rho_0} , 0 \right), &\\
		e^{\mu} \! _{\hat{2}} = &\left( \frac{L \cos \Psi}{\rho_0} , \frac{\sin \Psi}{U(\rho_0,0)} , \frac{E \cos \Psi}{\rho_0} , 0 \right), &\\
		e^{\mu} \! _{\hat{3}} = &\left( 0, 0, 0, \frac{1}{U(\rho_0,0)}  
		\right), &\\
	\end{aligned}
	\label{eq3circular}
\end{equation}
where 
\begin{equation} \label{psiphi}
  \Psi = \Psi(\phi) = \Psi_0 - \frac{E \rho_0^2 U^3(\rho_0,0)}{2L ^2 + \rho_0^2 U^2(\rho_0,0)} \phi.  
\end{equation}
The parameter $\Psi_0$, which is independent of $\phi$, represents the rotation angle between the tetrads $\{e_{ \hat{a}}\}$ and $\{\lambda_{ \hat{a}}\}$  when $\phi=0$. In particular, if we choose $\Psi_0=0$, then the tetrads will coincide when $\phi=0$. More generally, we can always set $\Psi_0$ so that the tetrads coincide at a particular angle $\phi$ (or, equivalently, at a particular proper time $\tau$).

\section{Tidal Forces in the Majumdar--Papapetrou Spacetime}

Tidal forces in General Relativity are associated with the geodesic deviation equation:
\begin{equation}
	\frac{D^{2}\xi^{\mu}}{D\tau^{2}} = R^{\mu} \!_{\nu \rho \delta}v^{\nu}v^{\rho}\xi^{\delta},
	\label{eq6}
\end{equation}
which determines the evolution of the geodesic deviation vector $\xi^{\mu}$ that connects infinitesimally close geodesics in terms of the Riemann tensor $R^{\mu} \!_{\nu \rho \delta}$ and the tangent vector $v^{\mu}(\tau)$ of the geodesic. In simple terms, a non-vanishing Riemann tensor indicates the existence of acceleration between neighboring geodesics and, consequently, the presence of tidal forces.

To derive the geodesic deviation equation, one chooses a one-parameter family of geodesics around a reference geodesic $x^\mu(\tau)$~\cite{dInverno:1992gxs,bookwald}. The family of geodesics defines a smooth two-dimensional surface $\widetilde{x}^\mu (s,\tau)$ on the spacetime in such a way that, for each $s=s_0$, the curve $\widetilde{x}^\mu (s_0,\tau)$ is a geodesic parametrized by the affine parameter $\tau$ with corresponding tangent vector $\widetilde{v}^\mu (s_0,\tau) = (\partial \widetilde x^\mu / \partial \tau) (s_0,\tau)$. The parameter $s$, defined in a neighborhood of $s=0$, parametrizes deviations from the reference geodesic $\widetilde{x}^\mu (0,\tau) = x^\mu(\tau)$. The deviation vector of a given geodesic in the family is defined by $\widetilde{\xi}^{\mu} (s_0,\tau) = (\partial \widetilde x^\mu / \partial s) (s_0,\tau) $. In particular, we denote the deviation vector $\widetilde{\xi}^{\mu} (0,\tau)$ of the reference geodesic as $\xi^{\mu} (\tau)$.
Since $(s,\tau)$ can be used as a coordinate system on the two-dimensional surface, the Lie derivative of the vector field $\widetilde{\xi}^{\mu}$ along $\widetilde{v}^\mu$ must vanish, meaning that the vector fields $\widetilde{\xi}^{\mu}$ and $\tilde{v}^\mu$ commute~\cite{dInverno:1992gxs,bookwald}:
\begin{equation} \label{firstorder}
\mathcal{L}_{\widetilde{v}}\widetilde{\xi}^{\mu} = [\widetilde{v},\widetilde{\xi}]^\mu = 0 \Rightarrow \widetilde{\xi}^{\nu} \nabla_\nu \widetilde{v}^{\mu} = \widetilde{v}^{\nu} \nabla_\nu \widetilde{\xi}^{\mu}.   
\end{equation}
By taking the derivative of the equation above along the direction of $\widetilde{v}^\mu$, one obtains the geodesic deviation equation after some algebraic manipulation.

Note that there is a critical difference between Equations~\eqref{eq6} and \eqref{firstorder}. While Equation~\eqref{eq6} is a second-order differential equation that is well defined independently of the choice of the one-parameter family of geodesics around $x^\mu(\tau)$, Equation~\eqref{firstorder} is a first-order differential equation that only makes sense if the family of geodesics is given {a priori}. This difference can be understood in the following sense: Equation~\eqref{eq6} has a larger space of solutions in comparison to Equation~\eqref{firstorder} since the arbitrariness of choosing the family of geodesics around $x^\mu (\tau)$ has not been used in Equation~\eqref{eq6}. In fact, the extra degrees of freedom in the initial conditions of Equation~\eqref{eq6} (in comparison to the initial conditions of  Equation~\eqref{firstorder}) correspond to this freedom of choosing the one-parameter family of geodesics. In the language of Riemannian and pseudo-Riemannian manifolds, the geodesic equation is typically referred to as the Jacobi equation. A key result in the mathematical literature is that every solution of the geodesic Equation~\eqref{eq6} corresponds to a one-parameter family of geodesics defined around $x^\mu (\tau)$ (see, e.g., Proposition 10.4 of Ref.~\cite{booklee}).

When analyzing geodesic deviations, it is convenient to apply the projection operator $h^{\mu} \!_{\nu} = \delta^{\mu} \!_{\nu}-v^{\mu}v_{\nu}$ on the deviation vector $\xi^{\mu}$ to obtain the so-called orthogonal connection vector $\eta^{\mu}$: 
\begin{equation}
	\eta^{\mu} = h^{\mu} \!_{\nu}\xi^{\nu}.
	\label{eq17}
\end{equation}
The components of the orthogonal connection vector in the tetrad basis,  $\eta^{\hat{a}}$, can be related to the components in the coordinate basis through
\begin{equation}
\eta^{\mu}= e^{\mu} \!\ _{\hat{a}} \eta^{\hat{a}}.
\end{equation}
An important characteristic of the orthogonal connection vector is the fact that the component along the direction of the associated geodesic vanishes, i.e., $\eta^{\hat{0}}=0$. With the help of the projection operator, the geodesic deviation equation in terms of $\eta^{\hat{a}}$ becomes~\cite{dInverno:1992gxs} 
\begin{equation}
	\frac{D^{2}\eta^{\hat{a}}}{D\tau^{2}} = R^{\mu} \!_{\nu \rho \delta} e^{\hat{a}} \!_{\mu} v^{\nu}v^{\rho}  e^{\delta} \!_{\hat{b}} \eta^{\hat{b}} =  R^{\hat{a}} \!_{\hat{0} \hat{0} \hat{b}} \eta^{\hat{b}} = K^{\hat{a}} \!_{\hat{b}} \eta^{\hat{b}} ,
	\label{eq7}
\end{equation}
where $K^{\hat{a}} \!_{\hat{b}}$ denotes the tidal tensor associated with the geodesic. Note that we can replace $D/D \tau$ in the equation above by $d / d \tau = e^{\mu} \ _{\! \! \hat{0}}\partial_\mu $.

To determine the explicit form of the tidal forces in the MP spacetime, we need the nonzero components of the Riemann tensor, which are (apart from the trivial symmetries $R_{\mu \nu \epsilon \delta} = R_{\epsilon \delta \mu \nu } = -R_{\nu \mu \epsilon \delta}$): 

\begin{equation}
\begin{aligned}
	R_{1010} =& \frac{1}{U^{4}}\left[ -\left(\frac{\partial U}{\partial z}\right)^{2} + 3\left(\frac{\partial U}{\partial \rho}\right)^{2} - U\left(\frac{\partial^{2} U}{\partial \rho^{2}}\right)\right],&\\
	R_{1030} =& \frac{1}{U^{4}}\left( 4\frac{\partial U}{\partial z} \frac{\partial U}{\partial \rho} - U\frac{\partial^{2} U}{\partial \rho \partial z} \right),&\\
	R_{2020} =& -\frac{1}{U^{4}}\left[ \rho^2 \left(\frac{\partial U}{\partial z}\right)^{2} + \rho U \frac{\partial U}{\partial \rho}  + \rho ^2  \left(\frac{\partial U}{\partial \rho} \right)^2 \right],	&\\
	R_{2121} =&  - \rho ^2 \left(\frac{\partial U}{\partial z}\right)^{2} + \rho ^2 \left(\frac{\partial U}{\partial \rho}\right)^{2} - \rho U  \frac{\partial U}{\partial \rho}  - \rho^{2} U \frac{\partial^{2} U}{\partial \rho^{2}} ,&\\
	R_{2132} =& -2 \rho^{2}  \frac{\partial U}{\partial z} \frac{\partial U}{\partial \rho} + \rho^{2}  U\frac{\partial^{2}U}{\partial \rho \partial z}, &\\
	R_{3030} =& \frac{1}{U^{4}}  \left[ 3\left( \frac{\partial U}{\partial z}	\right)^{2} - U\frac{\partial^{2} U}{\partial z^{2}} - \left( \frac{\partial U}{\partial \rho} \right)^{2} \right], &\\
	R_{3131} =& \left( \frac{\partial U}{\partial z} \right)^{2} + \left( \frac{\partial U}{\partial \rho} \right)^{2} - U \frac{\partial^{2} U}{\partial z^{2}} - U \frac{\partial^{2} U}{\partial \rho^{2}}, &\\
	R_{3232} =&   \rho^{2} \left( \frac{\partial U}{\partial z} \right)^{2} - \rho^{2} \left( \frac{\partial U}{\partial \rho} \right)^{2} - \rho^{2} U  \frac{\partial^{2}U}{\partial z^{2}} - \rho U \frac{\partial U}{\partial \rho}. &
	\end{aligned}
	\label{eq14}
\end{equation}

\subsection{Radial Geodesics}
For radial geodesics in the $z=0$ plane, we employ the freely falling frame given by \eqref{eq3} and find that the associated tidal tensor is diagonal. The non-trivial components of the geodesic deviation equation are given by
\begin{equation}
	\begin{aligned}
	\ddot{\eta}^{\hat{1}}= & \frac{1}{U^{4}} \left[U\frac{\partial^{2}U}{\partial \rho^{2}} - 3\left(\frac{\partial U}{\partial \rho}\right)^{2}\right]\eta^{\hat{1}}, 	&\\
	\ddot{\eta}^{\hat{2}} = & \frac{1}{U^{4}} \left[\left( \frac{\partial U}{\partial \rho}\right)^{2} - U\left( \frac{1}{\rho}\frac{\partial U}{\partial \rho} +  \frac{\partial^{2}U}{\partial \rho^{2}} \right)+E^{2}U^{3} \left( \frac{2}{\rho}\frac{\partial U}{\partial \rho}+\frac{\partial^{2} U}{\partial \rho^{2}}\right) \right]\eta^{\hat{2}}, &\\
	\ddot{\eta}^{\hat{3}} = &\frac{1}{U^{4}} \left[ \left( \frac{\partial U}{\partial \rho} \right)^{2} - U\left(\frac{\partial^{2}U}{\partial \rho^{2}}  + \frac{\partial^{2}U}{\partial z^{2}}  \right) + E^{2}U^{3} \left( \frac{\partial^{2}U}{\partial \rho^{2}} + 2\frac{\partial^{2}U}{\partial z^{2}} \right)
    \right]\eta^{\hat{3}}.&
	\end{aligned}
	\label{eq8}
\end{equation}

\begin{figure}[h]
\centering
\includegraphics[width=15.5cm]{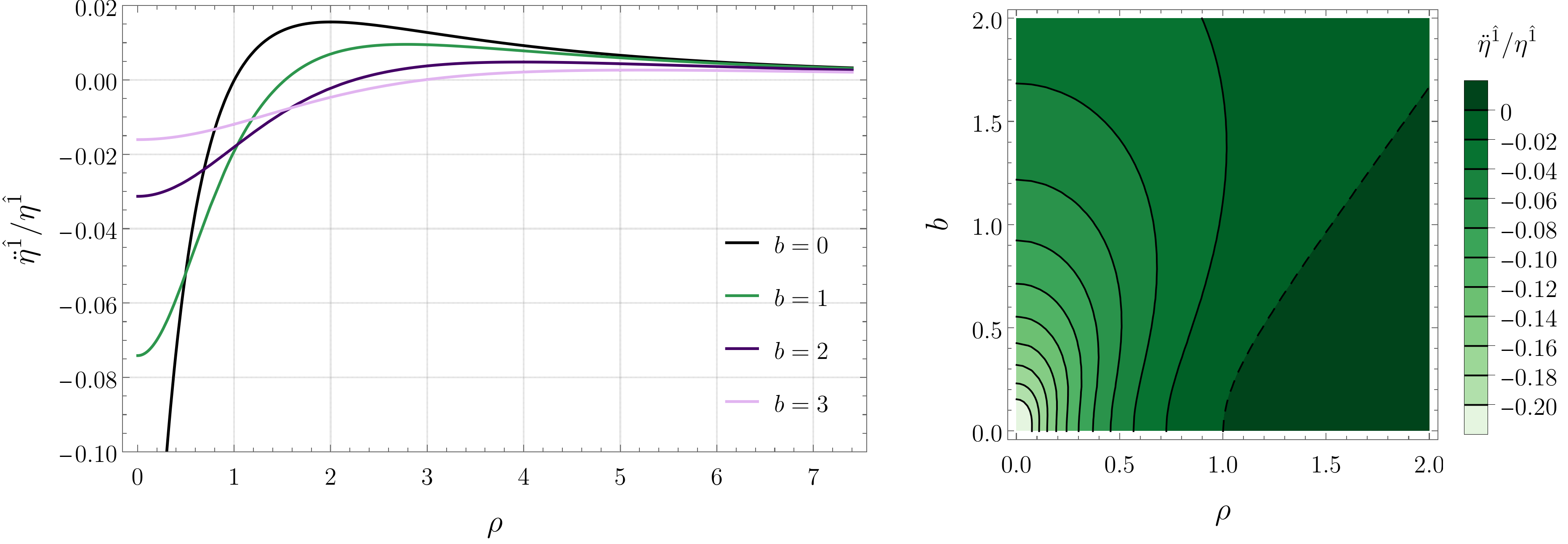}
\caption{Tidal force along the radial direction for radial timelike geodesics confined in the $z=0$ plane. (\textbf{Left}) Plots for several black hole separations $b$. (\textbf{Right}) Contour plot as a function of $b$ and $\rho$ (the dashed contour corresponds to the curve given by Equation~\eqref{radialzero} and indicates the transition from compression to stretching).} \label{fig1}
\end{figure}

\begin{figure}[h]
\centering
\includegraphics[width=15.5cm]{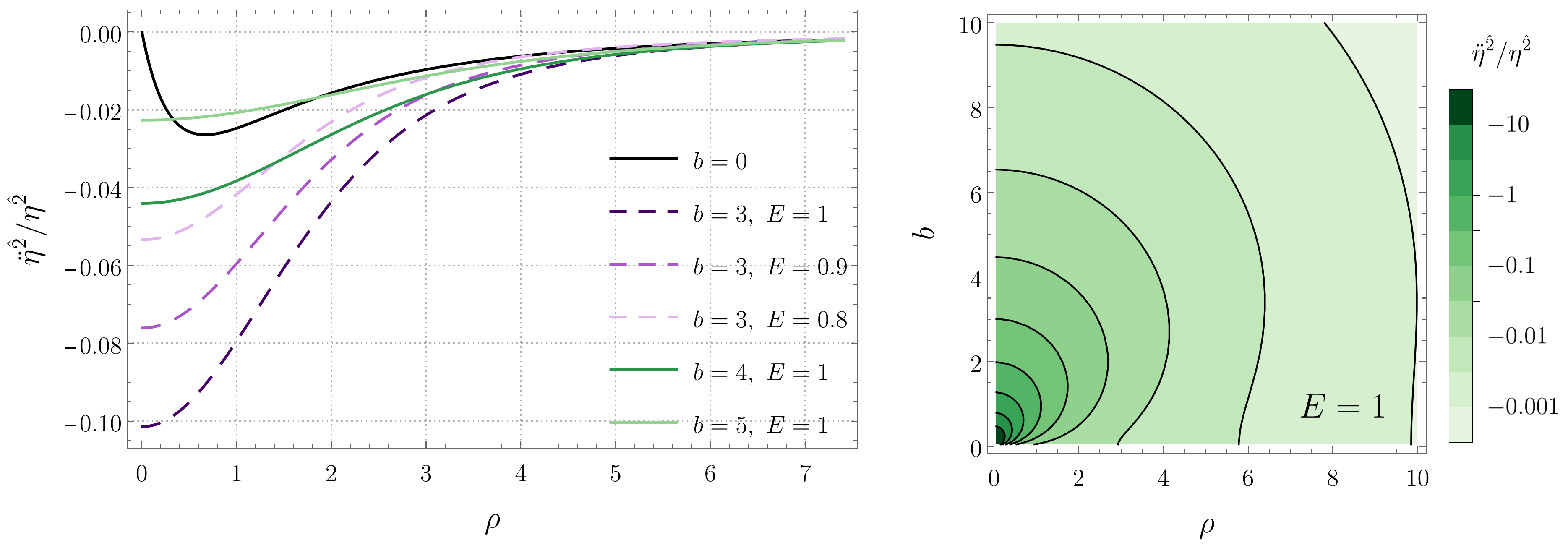}
\caption{Tidal force along the angular $\phi$ direction for radial timelike geodesics confined in the $z=0$ plane. (\textbf{Left}) Plots for several black hole separations $b$ and energies $E$. (\textbf{Right}) Contour plot as a function of $b$ and $\rho$ for $E=1$.} \label{fig2}
\end{figure}

\begin{figure}[h]
\centering
\includegraphics[width=15.5cm]{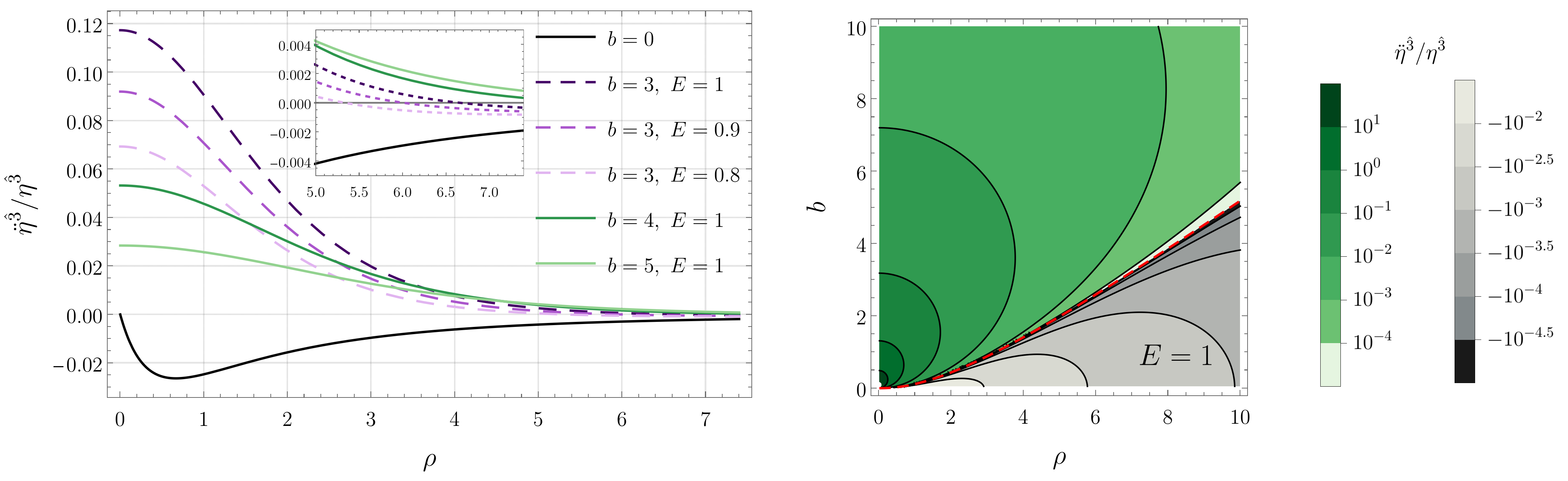}
\caption{Tidal force along the vertical $z$ direction for radial timelike geodesics confined in the $z=0$ plane. (\textbf{Left}) Plots for several black hole separations $b$ and energies $E$. (\textbf{Right}) Contour plot as a function of $b$ and $\rho$ for $E=1$ (the red dashed contour indicates the transition from compression to stretching).} \label{fig3}
\end{figure}

We remark that the potential $U$ and its derivatives in the expression above must be evaluated at $z=0$.
 The tidal forces in each direction correspond to the ratios between $\ddot \eta^{\hat{a}}$ and $\eta^{\hat{a}}$. The tidal forces along the radial ($\rho$), angular ($\phi$), and vertical ($z$) directions are shown, respectively, in Figures~\ref{fig1}--\ref{fig3}. For the sake of comparison, we have included in the left panels of each figure the corresponding tidal force when the MP metric describes an extremal Reissner--Nordstr\"om black hole (i.e.,~when $b=0$).


 Figure~\ref{fig1} tells us that, for sufficiently large values of the radial coordinate $\rho$, the radial tidal force is positive, meaning that it is a stretching force. As the origin $\rho=0$ is approached, the tidal force component attains a maximum value and then decreases, becoming negative (compression force) for sufficiently small values of $\rho$, attaining a minimum value at $\rho=0$. According to \eqref{eq8}, we see that the radial tidal force is independent of the energy $E$. The dependence on the separation $b$, however, is non-trivial. We note that increasing the separation $b$ between the black holes typically decreases the magnitude of the tidal forces at $z=0$. This is expected, since the black holes move away from the geodesics at $z=0$ when $b$ is increased. 
 
Unlike the radial tidal force, the angular force is always compressive for radial geodesics on the $z=0$ plane. This compressive force increases in magnitude with decreasing $\rho$. The magnitude of the compressive force also increases when the separation $b$ between the black holes decreases and when the energy $E$ associated with the geodesics increases.
Finally, the vertical component of the tidal force is compressive for large values of the radial coordinate $\rho$ and stretching for small values of $\rho$. The stretching force is most intense at the origin and, then, decreases monotonically with increasing $\rho$, changes sign, and reaches a local minimum, where the compressive force is most intense. As in the case of the angular component, the tidal force along the $z$ direction increases with the energy $E$ and decreases with the separation $b$.     

We note that the explicit form of Equation~\eqref{eq8}, after the substitution of the potential $U$ given in Equation~\eqref{eq2}, does not provide any useful analytical insights for the angular and vertical components. The expression for the radial component, on the other hand, is sufficiently simple, allowing one to determine an analytical expression of the coordinate where the tidal force vanishes. Explicitly, one has 
\begin{equation}
\ddot{\eta}^{\hat{1}}=  \frac{2M\left(2 \rho^2 - b^2 - 2M \sqrt{b^2 + \rho^2}\right)}{\sqrt{b^2 + \rho^2}\left(2M + \sqrt{b^2 + \rho^2} \right)^4} \,  \eta^{\hat{1}},
\end{equation}
meaning that the radial tidal force vanishes when
\begin{equation}
\rho = \sqrt{\frac{M^2 + b^2 + \sqrt{M^2 + 6 b^2}}{2}}. \label{radialzero}
\end{equation}

\subsection{Circular Geodesics}
For circular geodesics in the $z=0$ plane, we employ the freely falling frame given by \eqref{eq3circular} and find that the corresponding tidal tensor mixes the radial and angular components. The non-trivial components of the geodesic deviation Equation~\eqref{eq7} are
\begin{equation}
\begin{bmatrix}
\ddot{\eta}^{\hat{1}}\\
\ddot{\eta}^{\hat{2}} \\
\ddot{\eta}^{\hat{3}} 
\end{bmatrix} 
=
\big[\,\overline{K}\,\big]
\begin{bmatrix}
\eta^{\hat{1}}\\
\eta^{\hat{2}} \\
\eta^{\hat{3}} 
\end{bmatrix} = 
\begin{bmatrix}
\kappa_1 + \alpha \sin^2{\Psi} & \beta \sin { 2 \Psi} & 0\\
\beta \sin { 2 \Psi} & \kappa_2 - \alpha \sin^2{\Psi} & 0\\
0 & 0 & \kappa_3
\end{bmatrix}
\begin{bmatrix}
\eta^{\hat{1}}\\
\eta^{\hat{2}} \\
\eta^{\hat{3}} 
\end{bmatrix} 
	\label{eq8circular}
\end{equation}
The coefficients $\kappa_1$, $\kappa_2$, $\kappa_3$, $\alpha$, and $\beta$ of the matrix $\big[\,\overline{K}\,\big]$, which are constant along the circular geodesic, are given by
\begin{equation}
	\begin{aligned} 
	\kappa_1= & - \frac{2 L^4 (2\rho _0^2 U^2+3 L^2)}{\rho _0^4 U^4 \left(\rho _0^2 U^2+2 L^2\right)^2} + \left(\frac{\rho _0^2 U^2+2 L^2}{\rho _0^2 U^5} \right)\frac{\partial^{2}U}{\partial \rho^{2}}, &\\ 
\kappa_2= & - \frac{L^2 \left(\rho _0^2 U^2+ L^2\right)}{ \rho _0^2 U^2   \left(\rho _0^2 U^2+2 L^2\right)^2}, &\\ 
\kappa_3= & \left( \frac{ \rho _0^2 U^2+2 L^2}{\rho _0^2 U^5} \right) \frac{\partial^{2}U}{\partial z^{2}}, &\\ 
\alpha= & - \frac{ L^2 (\rho _0^4 U^4-3L^2 \rho _0^2 U^2 -6 L^4)}{ \rho _0^4 U^4 \left(\rho _0^2 U^2+2 L^2\right)^2} - \left(\frac{\rho _0^2 U^2+2 L^2}{\rho _0^2 U^5} \right)\frac{\partial^{2}U}{\partial \rho^{2}}, &\\ 
\beta= & - \frac{L^2 (\rho _0^4 U^4 - 3L^2 \rho _0^2 U^2 - 6 L^4 )}{2 \rho _0^4 U^4 \left(\rho _0^2 U^2+2 L^2\right)^2} - \left(\frac{\rho _0^2 U^2+2 L^2}{2 \rho _0^2 U^5} \right)\frac{\partial^{2}U}{\partial \rho^{2}}. & 
    \end{aligned}
	\label{compsKab}
\end{equation}

We remark that the potential $U$ and its derivatives in \eqref{compsKab} must be evaluated at $(\rho,z)=(\rho_0,0)$. According to \eqref{eq8circular}, the component of the tidal force in the $z$ direction is always decoupled from the other components, while the components in the other directions are coupled and oscillate as $\Psi=\Psi(\phi)$ changes along the geodesic. The tidal forces decouple when $\Psi=\Psi(\phi)$ is an integer multiple of $\pi$.

As mentioned previously, we have the freedom to align the tetrads $\{e_{ \hat{a}}\}$ and $\{\lambda_{ \hat{a}}\}$ at a specific point of the circular geodesic. In other words, we can choose $\Psi_0$ so that $\Psi=\Psi(\phi)$ vanishes at a specific angle $\phi$ of the circular geodesic (or, equivalently, at a specific proper time $\tau$). If we do that, then the tidal tensor becomes diagonal at that specific angle. For instance, if we set $\phi_0=0$ in \eqref{phievol} and choose $\Psi_0=0$ in \eqref{psiphi}, then the tidal forces in the radial and angular directions decouple at time $\tau=0$. In fact, whenever $\Psi$ is an integer multiple of $2\pi$, the tidal forces decouple according to:
%%%
\begin{equation}
	\ddot{\eta}^{\hat{1}}=  \kappa_1 \eta^{\hat{1}}, 	\qquad 
	\ddot{\eta}^{\hat{2}} =  \kappa_2 \eta^{\hat{2}}, \qquad
	\ddot{\eta}^{\hat{3}} =  \kappa_3 \eta^{\hat{3}}.
	\label{eq8circular2}
\end{equation}

\begin{figure}[h]
\centering
\includegraphics[width=15.5cm]{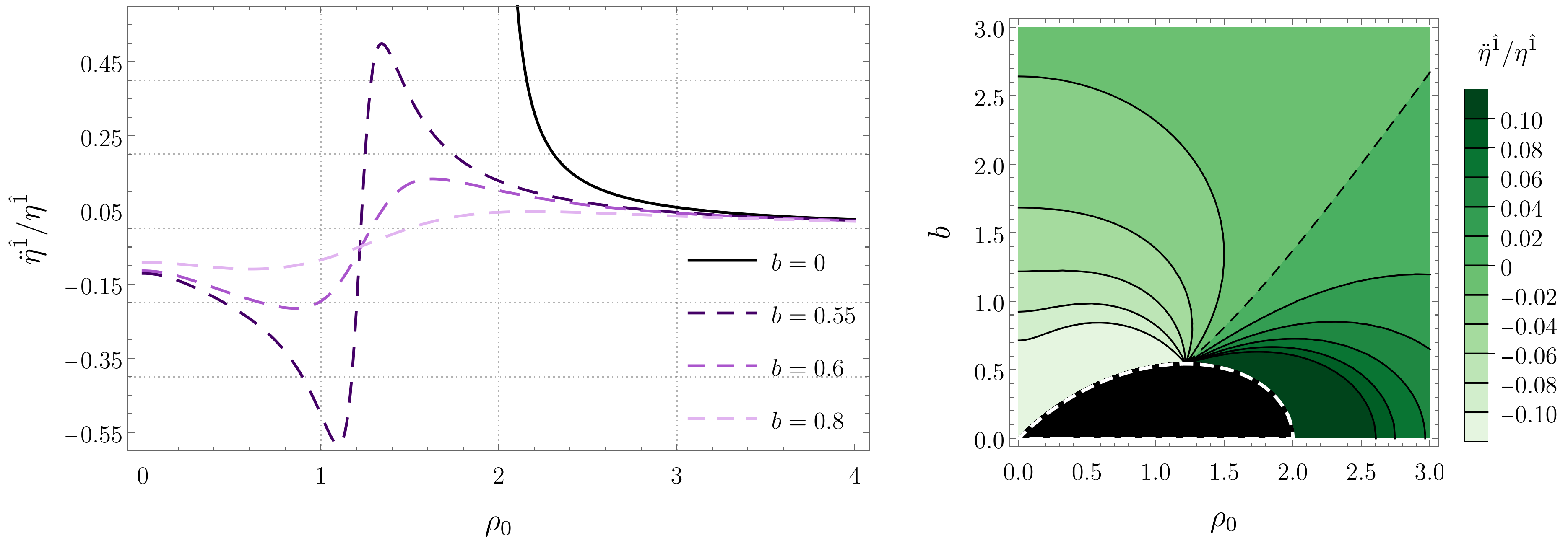}
\caption{Tidal force along the radial direction for circular timelike geodesics confined in the $z=0$ plane when $\Psi$ is an integer multiple of $2\pi$. (\textbf{Left}) Plots for several black hole separations $b$ as a function of the radius $\rho_0$ of the geodesic. (\textbf{Right}) Contour plot as a function of $b$ and $\rho_0$ (the dashed black contour indicates the transition from compression to stretching).} \label{fig5}
\end{figure}

\begin{figure}[h]
\centering
\includegraphics[width=15.5cm]{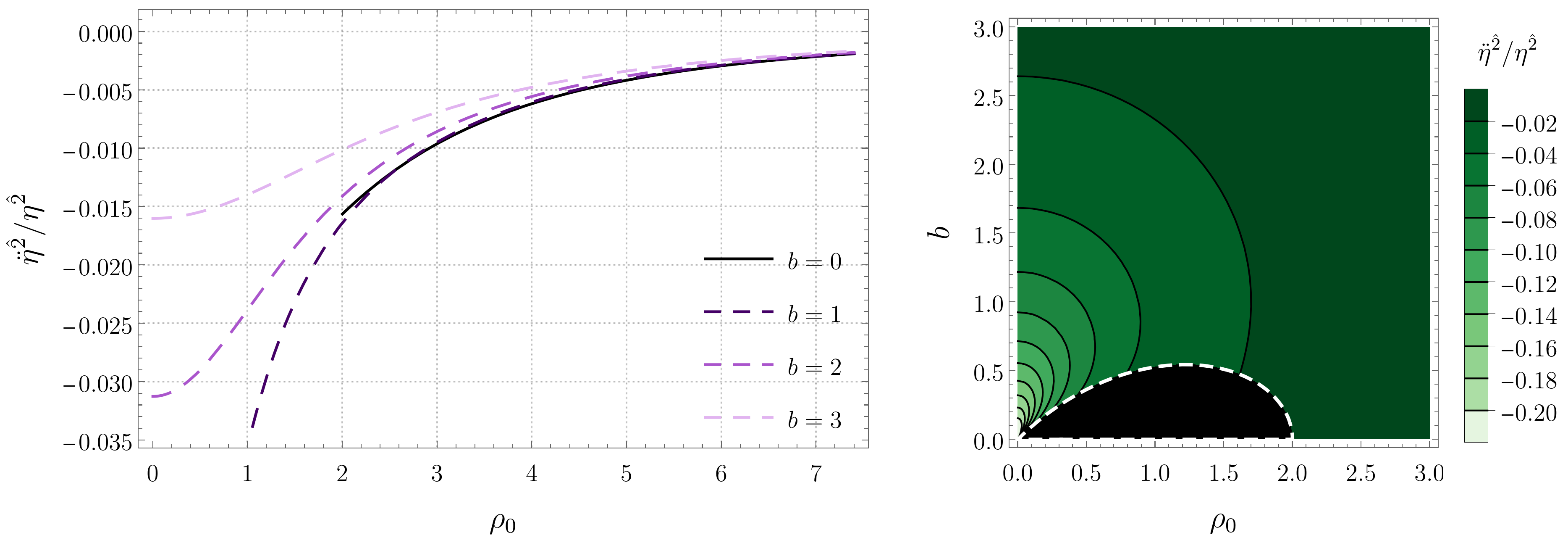}
\caption{Tidal force along the angular $\phi$ direction for circular timelike geodesics confined in the $z=0$ plane when $\Psi$ is an integer multiple of $2\pi$. (\textbf{Left}) Plots for several black hole separations $b$ as a function of the radius $\rho_0$ of the geodesic. (\textbf{Right}) Contour plot as a function of $b$ and $\rho_0$.} \label{fig6}
\end{figure}

\begin{figure}[h]
\centering
\includegraphics[width=15.5cm]{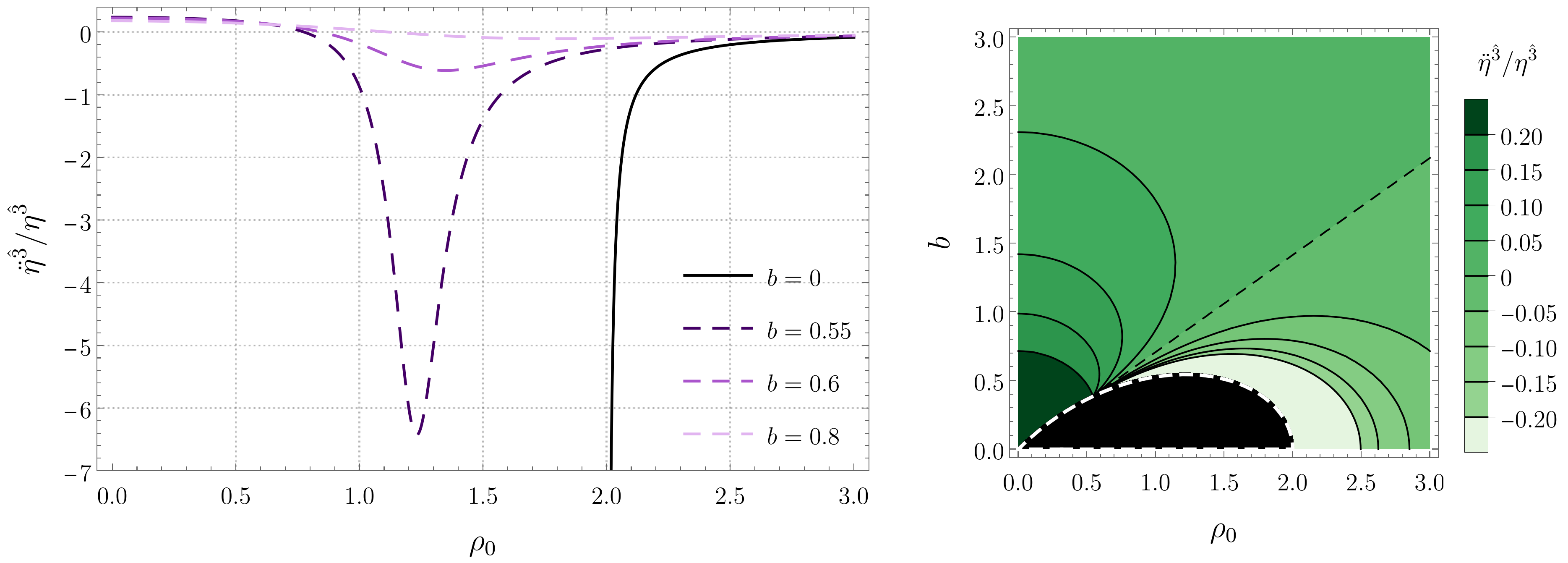}
\caption{Tidal force along the vertical $z$ direction for circular timelike geodesics confined in the $z=0$ plane when $\Psi$ is an integer multiple of $2\pi$. (\textbf{Left}) Plots for several black hole separations $b$ as a function of the radius $\rho_0$ of the geodesic. (\textbf{Right}) Contour plot as a function of $b$ and $\rho_0$ (the dashed black contour, which indicates the transition from compression to stretching, corresponds to the curve given by Equation~\eqref{radialzero}).} \label{fig7}
\end{figure}

Hence, at the points where $\Psi$ is an integer multiple of $2\pi$, the tidal forces along the directions $e_{ \hat{1}}$, $e_{ \hat{2}}$, and $e_{ \hat{3}}$ are given, respectively, by $\kappa_1$, $\kappa_2$, and $\kappa_3$. We plot such tidal forces as a function of the radius $\rho_0$ of the circular geodesic in Figures~\ref{fig5}--\ref{fig7}. For the sake of comparison, we have included in the left panels of each figure the corresponding tidal force when the MP metric describes an extremal Reissner--Nordstr\"om black hole (i.e.,~when $b=0$). These figures tell us how the tidal forces change when we move from one circular geodesic to another. In contrast, Figures~\ref{fig1}--\ref{fig3} exhibit how the tidal forces change along a given radial geodesic. Note that tidal forces associated with radial geodesics do not change as we move from one radial geodesic to another while keeping the distance $\rho$ constant (since Equation~\eqref{eq8} does not depend on the angle $\phi$).

We see in Figure \ref{fig5} that the tidal force in the radial direction is stretching for circular geodesics of sufficiently large radius, while for small radii, it is compressive. On the other hand, according to Figure \ref{fig7}, the behavior of the vertical tidal force is the opposite: stretching for small radii and compressive for large radii. From the explicit form of the potential $U$, given in Equation~\eqref{eq2}, we find that the vertical tidal force vanishes when the radius of the geodesic is 
\begin{equation}
\rho_0 = \sqrt{2} b. \label{verticalzero}
\end{equation}

The angular force shown in Figure \ref{fig6}, in contrast, is always compressive. We have also observed that, as the white dashed line in the right panels of Figures~\ref{fig5}--\ref{fig7} is approached, the radial and vertical components of the tidal force diverge while the angular component remains finite.


We close this section by noting that the eigenvalues of the matrix $\big[\,\overline{K}\,\big]$ are exactly $\kappa_1$, $\kappa_2$, and $\kappa_3$. The corresponding eigenvectors are, respectively, 
\begin{equation}
\big[w_1\big] = \begin{bmatrix}
\phantom{-}\cos \Psi \\
-\sin \Psi \\
\phantom{-}0 
\end{bmatrix}, \qquad
\big[w_2\big] = \begin{bmatrix}
\sin \Psi \\
\cos \Psi \\
0 
\end{bmatrix}, \qquad
\big[w_3\big] = \begin{bmatrix}
0 \\
0 \\
1 
\end{bmatrix} 
\end{equation}
Consequently, if we rotate back the vectors $e^{\mu} \! _{\hat{1}}$ and $e^{\mu} \! _{\hat{2}}$ by the angle $\Psi$, we obtain a new basis where the matrix $\big[\,\overline{K}\,\big]$ is diagonal. In other words, if we rotate the tetrad $\{e_{ \hat{a}}\}$ back to $\{\lambda_{ \hat{a}}\}$ at every point along the geodesic (consequently, undoing the rotation described in \eqref{e12mun}), the tensor $K^{\hat{a}} \!_{\hat{b}}$ becomes diagonal. However, since $\{\lambda_{ \hat{a}}\}$ is not parallel propagated along the geodesic (i.e.,~it is not a freely falling frame), the left-hand side of Equation~\eqref{eq8circular} would not be the second derivative of the rotated vectors.

\section{Numerical Solutions of the Geodesic Deviation Equation}

We can also investigate tidal effects by determining the components of the deviation vector $\eta^{\hat{a}}$. We proceed as in \cite{paper01,paper02,paper05,Sharif:2018gzj,Shahzad:2017vwi,Hong:2020bdb,Li:2021izh,Vandeev:2021yan,Uniyal:2022ouc} by directly solving the geodesic deviation Equation~\eqref{eq6} with appropriate initial conditions. Being a second-order differential equation, the initial-value problem for Equation~\eqref{eq6} requires an initial condition for both the deviation vector and its first-order derivative.
We impose boundary conditions at the starting time $\tau=\tau_0$, corresponding to the spacetime event $x_0=x(\tau_0)=(t_0,\rho_0,\phi_0,0)$. In our calculations, we have set $\rho_0 = 10$ and $\phi_0=0$ (we also set $\Psi_0=0$ in Equation~\eqref{psiphi} and leave $t_0$ unspecified since the results do not depend on it). We consider two different initial conditions (denoted by IC-I and IC-II) to solve the geodesic deviation equation:
\begin{align}
		\text{IC-I}: \ &\eta^{\hat{a}}(\tau_0) =1, \qquad  \dot{\eta}^{\hat{a}}(\tau_0)=0; \label{ic-i} \\
		\text{IC-II}: \ & \eta^{\hat{a}}(\tau_0) =0, \qquad \dot{\eta}^{\hat{a}}(\tau_0)=1.  \label{ic-ii}  
	\end{align}
Initial condition IC-I can be understood as representing the deviation between two geodesics that are initially parallel (in the sense that the rate of change of the their deviation vanishes at $\tau=\tau_0$). Initial condition IC-II, on the other hand, corresponds to two initially diverging geodesics that intercept at the initial time $\tau_0$ (at the point $x^{\mu}(\tau_0)$).

\subsection{Radial Geodesics}

When analyzing radial geodesics, it is convenient to replace derivatives with respect to the proper time $\tau$ by derivatives with respect to the radial coordinate $\rho$. To accomplish this, we use the relation:  
\begin{equation}
		\frac{d}{d\tau} = \frac{d\rho}{d\tau}\frac{d}{d\rho}=-\sqrt{E^{2}--\frac{1}{U^{2}(\rho,0)}}\frac{d}{d\rho}, 
	\label{eq25}
\end{equation}
which follows from Equation~\eqref{eq11}, to transform Equation~\eqref{eq8} into:
\begin{align}
  \frac{d}{d\rho} \left[\frac{\left(E^{2}U^{2}(\rho,0)- 1\right)^{3/2}}{U^{3}(\rho,0)}\frac{d}{d\rho} \left(  \frac{U(\rho,0) \eta^{\hat{1}} }{\sqrt{E^{2}U^{2}(\rho,0)-1}} \right)\right] &= 0,   \label{eq26i} \\
  \frac{d}{d\rho} \left[ \rho^2 U (\rho,0) \sqrt{E^{2}U^{2}(\rho,0) - 1} \ \frac{d}{d\rho} \left(  \frac{\eta^{\hat{2}}}{\rho  U (\rho,0)} \right) \right] &= 0, \label{eq26ii}\\
	\frac{d}{d\rho} \left[U (\rho,0) \sqrt{E^{2}U^{2}(\rho,0)-1} \frac{d}{d\rho} \left(  \frac{\eta^{\hat{3}}}{U (\rho,0)} \right) \right] + \frac{2 E^2 U^2 (\rho,0)-1}{\sqrt{E^{2}U^{2}(\rho,0)-1}} \frac{\frac{\partial ^2 U(\rho,0)}{\partial z^2}}{U (\rho,0)} \eta^{\hat{3}} &= 0.
	\label{eq26iii}
 \end{align}

The equations for the radial and the angular components of the geodesic deviation can be integrated by quadratures:
\begin{align}
\eta^{\hat{1}}(\rho)  &= \frac{\sqrt{E^{2}U^{2}(\rho,0)-1}}{U(\rho,0)}\left(K_1 + K_2 \int_{\rho_0} ^{\rho} \frac{U^3(\rho',0)}{\left(E^{2}U^{2}(\rho',0)- 1\right)^{3/2}} \, d\rho' \right), \label{eta1quadrature}\\
\eta^{\hat{2}}(\rho)  &= \rho U(\rho,0)\left(K_3 + K_4 \int_{\rho_0} ^{\rho} \frac{d\rho'}{\rho'{}^2 U(\rho',0)\sqrt{E^{2}U^{2}(\rho',0)-1}} \, d\rho' \right),
\label{eta2quadrature}
\end{align}
%%%
where $K_1$, $K_2$, $K_3$, and $K_4$ are constants of integration that can be associated with the initial conditions for the differential equations. The component of the geodesic deviation along the $z$ direction, on the other hand, cannot be cast in terms of simple integrals.

\begin{figure}[h]
\centering
\includegraphics[width=15.5cm]{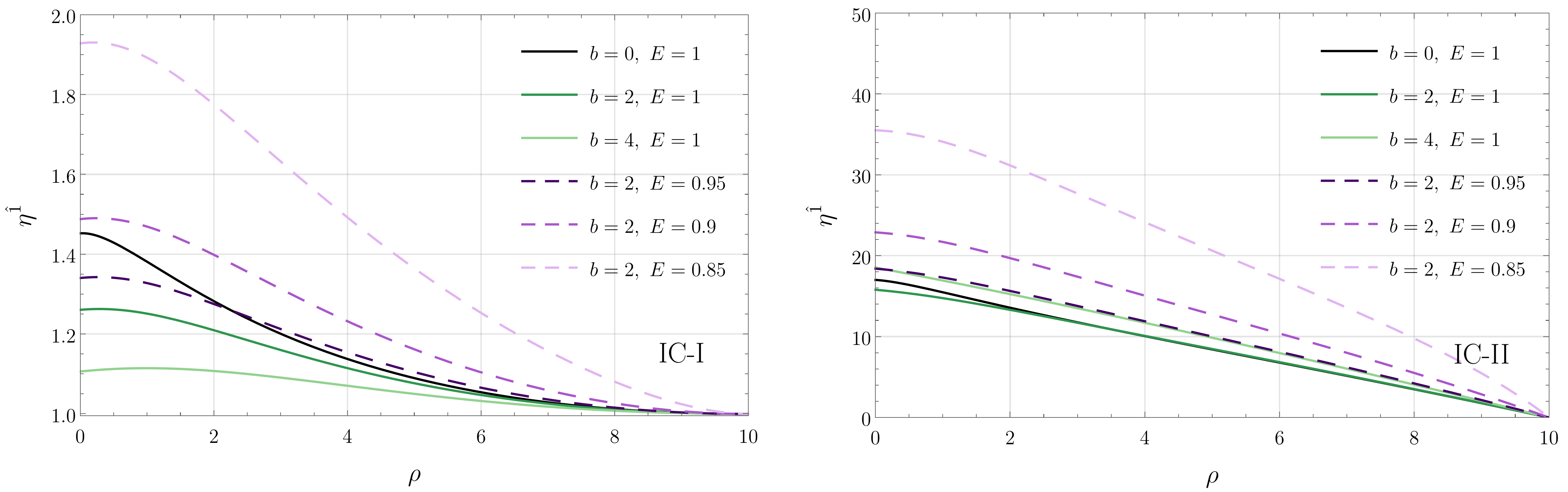}
\caption{Component $\eta^{\hat{1}}$ of the connection vector associated with radial geodesics in the $z=0$ plane as a function of $\rho$ for IC-I (\textbf{left}) and for IC-II (\textbf{right}). Each curve corresponds to a radial geodesic specified by the parameters $b$ and $E$ indicated in the panels.} \label{fig8}
\end{figure}

\begin{figure}[h]
\centering
\includegraphics[width=15.5cm]{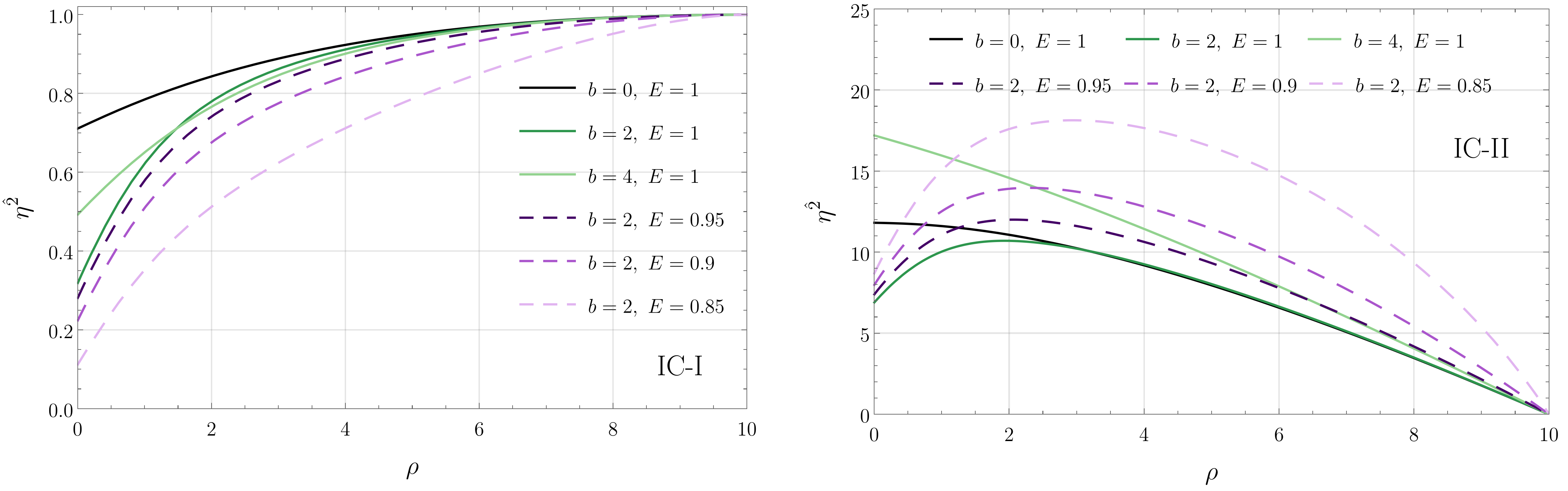}
\caption{Component $\eta^{\hat{2}}$ of the connection vector associated with radial geodesics in the $z=0$ plane as a function of $\rho$ for IC-I (\textbf{left}) and for IC-II (\textbf{right}). Each curve corresponds to a radial geodesic specified by the parameters $b$ and $E$ indicated in the panels.} \label{fig9}
\end{figure}

\begin{figure}[h]
\centering
\includegraphics[width=15.5cm]{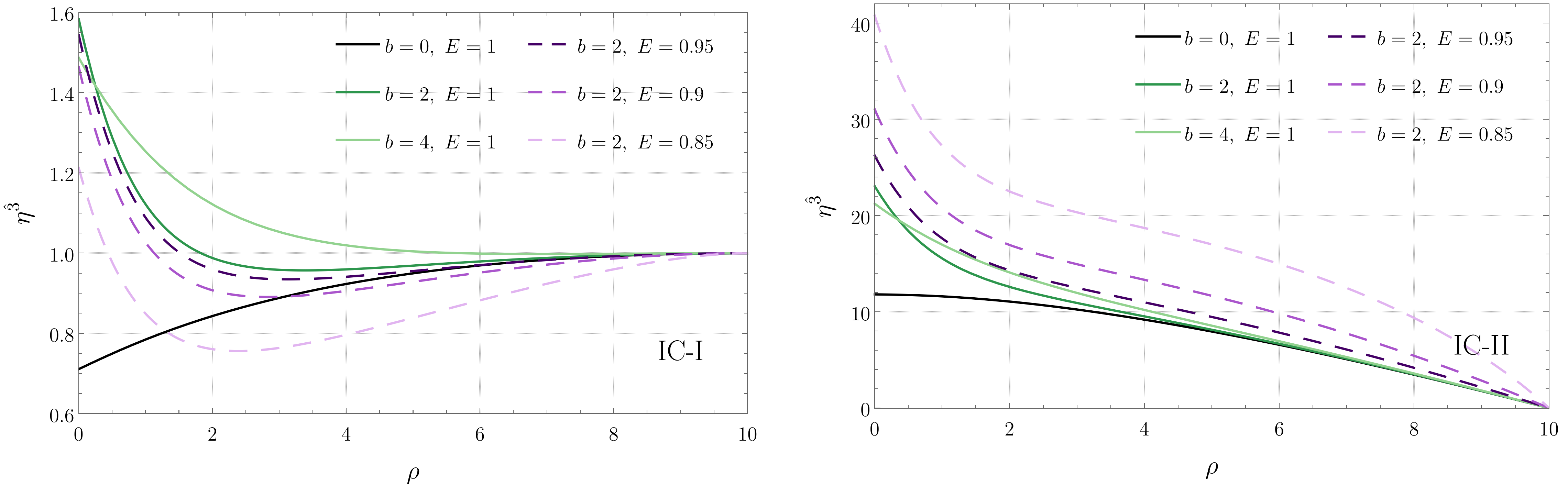}
\caption{Component $\eta^{\hat{3}}$ of the connection vector associated with radial geodesics in the $z=0$ plane as a function of $\rho$ for IC-I (\textbf{left}) and for IC-II (\textbf{right}). Each curve corresponds to a radial geodesic specified by the parameters $b$ and $E$ indicated in the panels.} \label{fig10}
\end{figure}

In order to solve the differential Equations \eqref{eq26i}--\eqref{eq26iii}, we impose the initial conditions \eqref{ic-i} and \eqref{ic-ii}. Note that, in terms of the radial coordinate, the initial condition for the derivative becomes: 
\begin{equation}
\frac{d \eta^{\hat{a}}}{d \rho}(\rho_0) = - \frac{ U(\rho_0,0)}{\sqrt{E^{2}U^{2}(\rho_0,0)-1}} \dot{\eta}^{\hat{a}}(\tau_0). 
\end{equation}
Additionally, due to Equation~\eqref{eq11}, the energy of the associated geodesics must satisfy the constraint $|E| \geq 1/U(\rho_0,0)$.
The numerical solution of Equations~\eqref{eq26i}--\eqref{eq26iii}, corresponding to the radial $\eta^{\hat{1}}(\rho)$, angular $\eta^{\hat{2}}(\rho)$, and vertical $\eta^{\hat{3}}(\rho)$  components of the orthogonal connection are shown in Figures \ref{fig8}--\ref{fig10}, respectively. In each plot, we indicate the separation $b$ between the black holes and the energy $E$ of the corresponding geodesic. The left panel in each figure corresponds to the initial condition IC-I, while the right panel corresponds to IC-II. We see that the radial deviation is largest at $\rho=0$ and decreases as one moves away from the origin for both IC-I and IC-II. On the other hand, the angular deviation for IC-I increases with $\rho$, while the qualitative behavior for IC-II depends on the choices of $b$ and $E$. Regarding the vertical $z$ direction, we see that, while the qualitative behavior of $\eta^{\hat{3}}$ is sensitive to the specific values of $b$ and $E$ for IC-I, it is monotonically decreasing with $\rho$ for IC-II.

Figure~\ref{fig8} also shows that, when the energy of the geodesic increases, the radial deviation typically decreases as well. The same behavior with respect to the energy is observed for the angular $\eta^{\hat{2}}(\rho)$ and vertical $\eta^{\hat{3}}(\rho)$ components of the orthogonal connection, shown, respectively, in Figures~\ref{fig9} and \ref{fig10}, when IC-II is considered. On the other hand, for IC-I, we see that an increase of the energy typically increases the deviation in the angular and vertical directions.

\subsection{Circular Geodesics}

When considering circular geodesics, it is convenient to replace the derivatives with respect to the proper time $\tau$ by the derivatives with respect to the angular coordinate $\phi$. To accomplish this, we use the relation:  
\begin{equation}
		\frac{d}{d\tau} = \frac{d\phi}{d\tau}\frac{d}{d\phi}=\frac{L}{\rho_0^2 U^2(\rho_0,0)}\frac{d}{d\phi}, 
	\label{eq25circular}
\end{equation}
which follows from Equation~\eqref{eq12}. As a consequence, the derivatives in the left-hand side of Equation~\eqref{eq8circular} are replaced by the derivatives with respect to the angle $\phi$, while the matrix $\big[\,\overline{K}\,\big]$   is multiplied by $\rho_0^4 U^4(\rho_0,0)/L^2$. Analogously, in terms of $\phi$, the initial condition for the derivatives in \eqref{ic-i} and \eqref{ic-ii} becomes: 
\begin{equation}
\frac{d \eta^{\hat{a}}}{d \phi}(\phi_0) = \frac{L}{\rho_0^2 U^2(\rho_0,0)} \dot{\eta}^{\hat{a}}(\tau_0). 
\end{equation}

\begin{figure}[h]
\centering
\includegraphics[width=15.5cm]{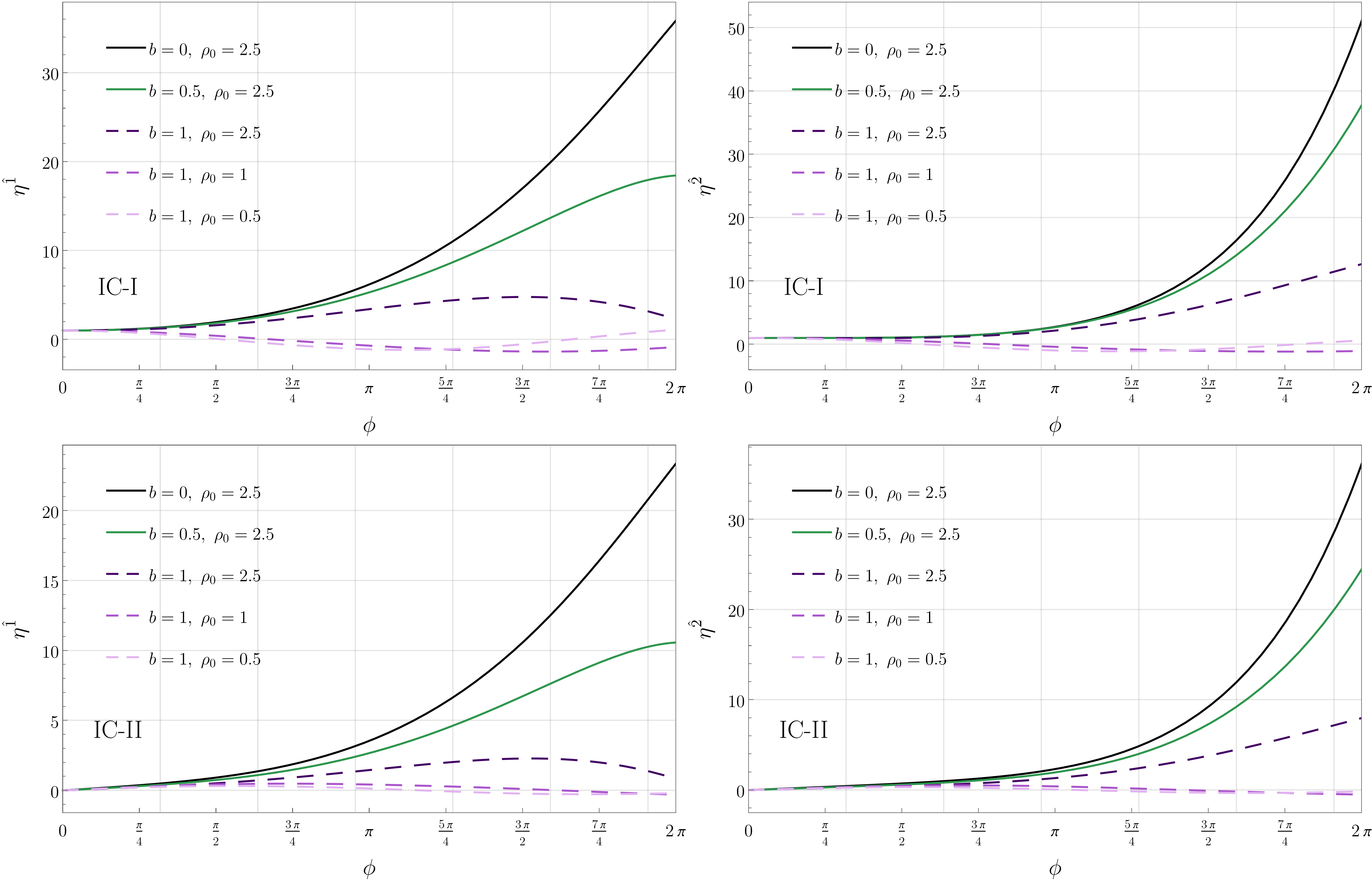}
\caption{Components $\eta^{\hat{1}}$ (\textbf{top}) and $\eta^{\hat{2}}$ (\textbf{bottom}) of the connection vector associated with circular geodesics in the $z=0$ plane as a function of $\phi$ for IC-I (\textbf{left}) and for IC-II (\textbf{right}). Each curve corresponds to a circular geodesic specified by the parameters $b$ and $\rho_0$ indicated in the panels.} \label{fig11}
\end{figure}

\begin{figure}[h]
\centering
\includegraphics[width=15.5cm]{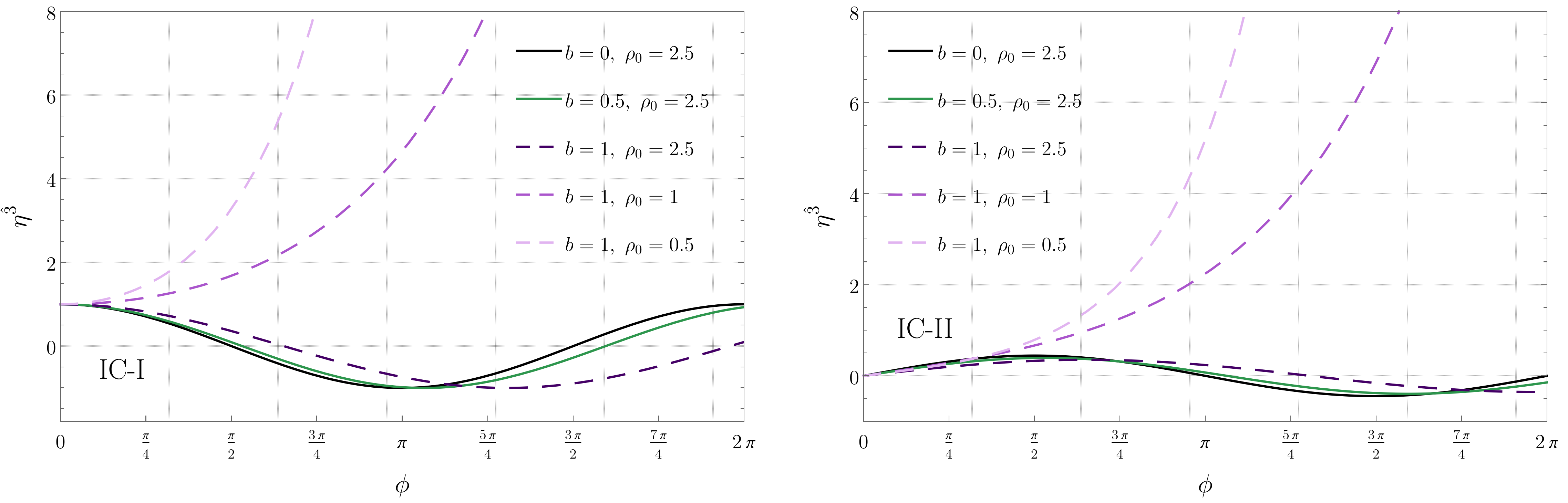}
\caption{Components $\eta^{\hat{3}}$ of the connection vector associated with circular geodesics in the $z=0$ plane as a function of $\phi$ for IC-I (\textbf{left}) and for IC-II (\textbf{right}). Each curve corresponds to a circular geodesic specified by the parameters $b$ and $\rho_0$ indicated in the panels.} \label{fig12}
\end{figure}

In Figures \ref{fig11} and \ref{fig12}, we show the numerical solutions of Equation~\eqref{eq8circular} for both initial conditions IC-I (left panels) and IC-II (right panels).\footnote{Since $\eta^{\hat{1}}$ and $\eta^{\hat{2}}$ are coupled through Equation~\eqref{eq8circular}, to avoid ambiguities, we specify that IC-I (similarly, IC-II) means that the condition \eqref{ic-i} (similarly, \eqref{ic-ii}) is imposed simultaneously on $\eta^{\hat{1}}$ and $\eta^{\hat{2}}$.} In each panel, we indicate the separation $b$ between the black holes and the radius of the corresponding circular geodesics (the associated values of energy $E$ and angular momentum $L$ are given in Figure \ref{fig0}).
We have observed that the geodesic deviation in the directions $e_{\hat{1}}$ and $e_{\hat{2}}$  will diverge with the angle $\phi$ unless $b$ is sufficiently large. Conversely, the deviation in the vertical direction $e_{\hat{3}}$ will diverge unless $b$ is sufficiently small.

Such a behavior can be understood analytically for the vertical component $\eta^{\hat{3}}$ (since $\eta^{\hat{1}}$ and $\eta^{\hat{2}}$ are coupled, a similar analysis for the radial and angular components is not possible). After plugging in the explicit form of $U$ into Equation~\eqref{eq8circular} and using Equation~\eqref{circreq1}, we find that the differential equation for $\eta^{\hat{3}}$ becomes
\begin{equation}
    \frac{d^2 \eta ^{\hat{3}}}{d \phi ^2} + \left(\frac{\rho_0^2 - 2 b^2}{\rho_0^2 + b^2} \right) \eta ^{\hat{3}} = 0.
\end{equation}

The corresponding solution is
\begin{equation}
    \eta ^{\hat{3}}(\phi) = K_5 \exp \left(\sqrt{\frac{2 b^2 - \rho_0^2 }{\rho_0^2 + b^2}} \, \phi \right) + K_6 \exp \left(-\sqrt{\frac{2 b^2 - \rho_0^2 }{\rho_0^2 + b^2}} \, \phi \right),
\end{equation}
where $K_5$ and $K_6$ are constants associated with the initial conditions. 
It is straightforward to see that the solutions will be exponentially growing if $b$ is sufficiently large ($b > \rho_0 / \sqrt{2}$) and oscillatory otherwise ($b < \rho_0 / \sqrt{2}$). In fact, considering that the vertical tidal force, shown in Figure \ref{fig7}, vanishes exactly when $b = \rho_0 / \sqrt{2}$, we conclude that the deviation vector in the vertical direction will be bounded if the tidal force in the vertical direction is compressive.

%%%%%%%%%%%%%%%%%%%%%%%%%%%%%%%%%%%%%%%%%%
\section{Discussion}

In this work, we investigated tidal forces acting on geodesics of the MP spacetime that describes two extremely charged black holes in equilibrium. We focused on radial geodesics (vanishing angular momentum $L=0$) and circular geodesics (constant $\rho=\rho_0$) that live on the plane equidistant to two black holes of equal masses. In Section IV, we  analyzed the radial, angular, and vertical components of the tidal forces, while in Section V, we analyzed the numerical solutions of the geodesic deviation equation for two classes of initial conditions. 

In particular, we have seen that the tidal forces in the radial direction, for both radial and circular geodesics, can be either compressive or stretching: near the origin, they are compressive, and far away, they are stretching. A similar behavior has been observed for Reissner--Nordstr\"om black holes in Ref.~\cite{paper01}. However, unlike Reissner--Nordstr\"om black holes, the tidal forces in the angular direction are always compressive for the geodesics of the MP that were analyzed. The tidal forces in the vertical direction, as in the case of the radial component, exhibit a sign change with respect to the distance. Nevertheless, the qualitative behavior is quite the opposite of their radial counterparts: for small radii, the force is stretching, and for large radii, it is compressive.  Finally, we solved numerically the differential equations that determine the geodesic deviation vector as a function of the radial distance. We employed two types of initial conditions and analyzed the behavior of the corresponding solutions. 

As a future research direction, one could investigate tidal forces in a binary system of rotating black holes~\cite{Cabrera-Munguia:2018omi,Manko:2018iyn,Ramirez-Valdez:2020rak,Baez:2022dqu,Camilloni:2023rra,Camilloni:2023xvf} or in deformed Kerr spacetimes~\cite{Konoplya:2016pmh,Franzin:2021kvj,Siqueira:2022tbc}. Another interesting line of investigation would be to consider geodesics in analog black hole spacetimes~\cite{Barcelo:2005fc,Jacquet:2020bar}. In particular, one possibility would be to explore analog tidal effects in the double-sink solution of hydrodynamics~\cite{Assumpcao:2018bka} or in vortices that form in condensates~\cite{Garay:1999sk,Jacquet:2020znq,Giacomelli:2019tvr,Patrick:2021oqk,Patrick:2021dxw,Cardoso:2022yin,Solnyshkov:2023sjl}.

%%%%%%%%%%%%%%%%%%%%%%%%%%%%%%%%%%%%%%%%%%%%%%%%%%%%%%%%%%%%%%%%%%%%%%%%%%%%%%%%%%%%%%%%%%%%%%%%

\acknowledgments

The authors would like to thank Haroldo C.~D.~Lima Junior and Luís C.~B.~Crispino for enlightening discussions. This research was partially financed by the Coordenação de Aperfeiçoamento de Pessoal de Nível Superior (CAPES, Brazil)---Finance Code 001. M.R. also acknowledges partial support from the Conselho Nacional de Desenvolvimento
Científico e Tecnológico (CNPq, Brazil), Grant No. FA 315664/2020-7, and from The São Paulo Research
Foundation (FAPESP, Brazil), Grant No. 2022/08335-0. \\

\textbf{Supplemental material:} a Mathematica notebook, which can be used to reproduce the plots of the article, is provided in~\cite{albacete_2024_10607824}.

%%%%%%%%%%%%%%%%%%%%%%%%%%%%%%%%%%%%%%%%%%%%%%%%%%%%%%%%%%%%%%%%%%%%%%%%%%%%%%%%%%%%%%%%%%%%%%%%

%\bibliographystyle{apsrev4-2}
\bibliography{tidalforces_MP.bbl}

\end{document}